# Spin Hall Magnetoresistance in Antiferromagnetic Insulators

Stephan Geprägs[1], Matthias Opel[1], Johanna Fischer[1,2*], Philipp Schwenke[1,2+], Matthias Althammer[1,2], Hans Huebl[1,2,3], and Rudolf Gross[1,2,3]

[1]Walther-Meissner-Institut, Bayerische Akademie der Wissenschaften, 85748 Garching, Germany
[2]TUM School of Natural Sciences, Technische Universität München, 85748 Garching, Germany
[3]Munich Center for Quantum Science and Technology (MCQST), 80799 München, Germany
*present address: SPINTEC, 38054 Grenoble Cedex 9, France
[+]present address: RPTU Kaiserslautern Landau, 67663 Kaiserslautern, Germany

**Antiferromagnetic materials promise improved performance for spintronic applications, as they are robust against external magnetic field perturbations and allow for faster magnetization dynamics compared to ferromagnets. The direct observation of the antiferromagnetic state, however, is challenging due to the absence of a macroscopic magnetization. We show that the spin Hall magnetoresistance (SMR) is a versatile tool to probe the antiferromagnetic spin structure via simple electrical transport experiments by investigating the easy-plane antiferromagnetic insulators α-$Fe_2O_3$ (hematite) and NiO in bilayer heterostructures with a Pt heavy-metal top electrode. While rotating an external magnetic field in three orthogonal planes, we record the longitudinal and the transverse resistivities of Pt and observe characteristic resistivity modulations consistent with the SMR effect. We analyze both their amplitude and phase and compare the data to the results from a prototypical collinear ferrimagnetic $Y_3Fe_5O_{12}$/Pt bilayer. The observed magnetic field dependence is explained in a comprehensive model, based on two magnetic sublattices and taking into account magnetic field-induced modifications of the domain structure. Our results show that the SMR effect allows to readout the spin configuration and to investigate magnetoelastic effects in antiferromagnetic multi-domain materials. We demonstrate that the SMR amplitude scales with the sum of the absolute sublattice magnetizations in ferrimagnetic and antiferromagnetic materials. In α-$Fe_2O_3$/Pt bilayers, we find an unexpectedly large SMR amplitude of 2.5 x $10^{-3}$, twice as high as for prototype $Y_3Fe_5O_{12}$/Pt bilayers, making the system particularly interesting for room-temperature antiferromagnetic spintronic applications.**

## I. Introduction

Compared to ferromagnets, antiferromagnetic (AF) materials promise improved performance for spintronic devices: they are robust against external magnetic field perturbations and allow for faster magnetization dynamics. The correct determination of their AF state, however, is challenging due to the absence of a macroscopic magnetization. Recently [1], we demonstrated that the AF spin structure can be probed and readout via simple electrical transport experiments, utilizing the spin Hall magnetoresistance (SMR). Representing a well-known manifestation of spin current physics, the SMR effect is based on an interfacial exchange of angular momentum between localized magnetic moments in magnetically ordered insulators and conduction electrons in adjacent metallic top electrodes with large spin-orbit coupling [2,3].

## II. Experimental

We investigate the easy-plane antiferromagnetic insulators (AFI) α-$Fe_2O_3$ (hematite) [4] and NiO [5] in epitaxial thin film bilayer heterostructures with a heavy metal Pt top electrode. In angle-dependent magnetoresistance (ADMR) measurements, we rotate an external magnetic field in the film plane, record the longitudinal and the transverse resistivity of Pt, and observe characteristic resistivity modulations (blue and red in Fig. 1). The data follow a sinusoidal behavior, expected from SMR theory [3] and reported earlier for prototypical collinear ferrimagnetic insulating (FMI) $Y_3Fe_5O_{12}$ (YIG)/Pt bilayers (black in Fig. 1) [2].

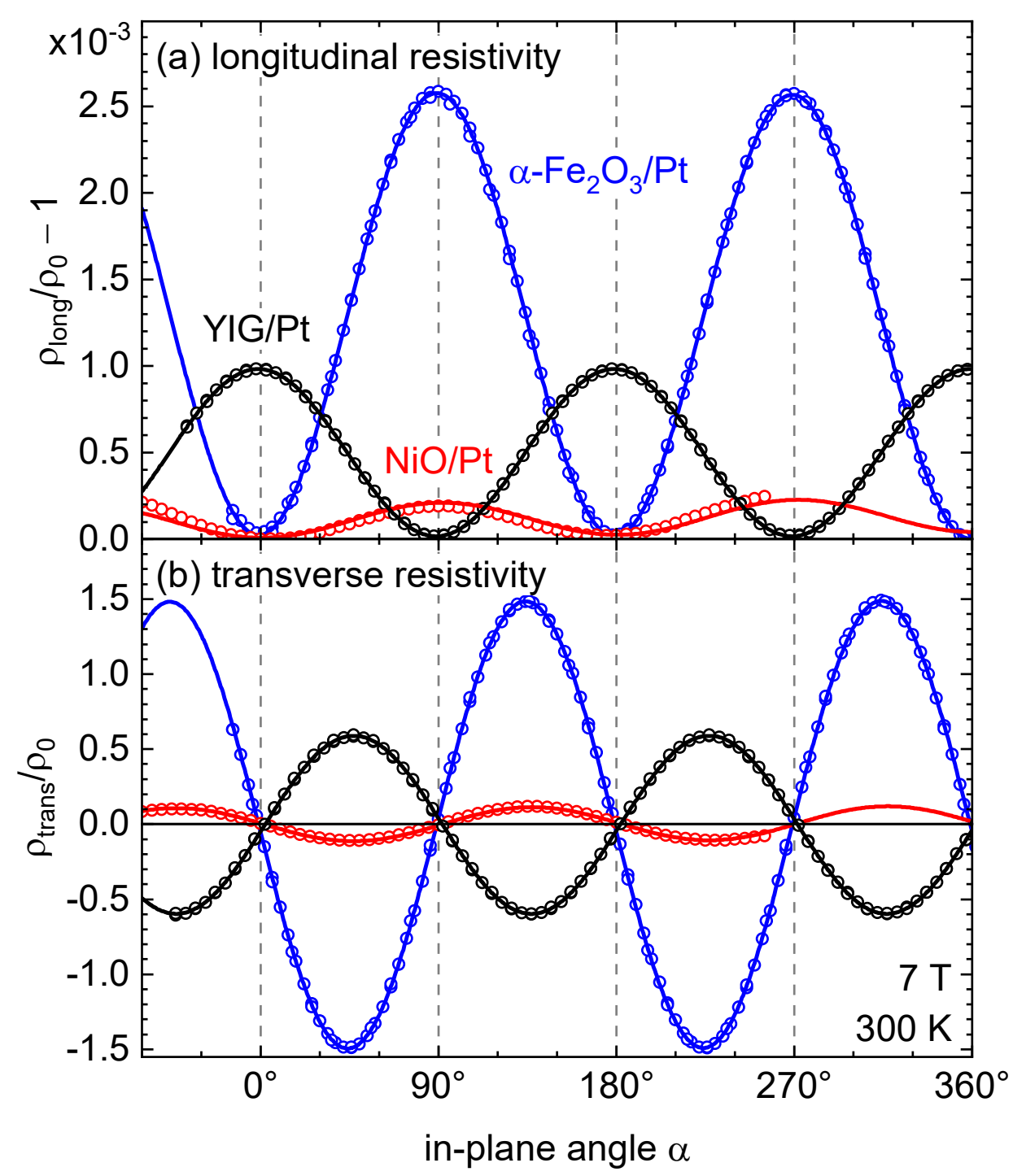

Fig. 1. In-plane ADMR at 300 K in a magnetic field of 7 T of antiferromagnetic (0001)-oriented α-$Fe_2O_3$/Pt (blue) and (111)-oriented NiO/Pt (red) bilayers as well as a ferrimagnetic (001)-oriented $Y_3Fe_5O_{12}$ (YIG)/Pt bilayer (black). The symbols represent the normalized (a) longitudinal ($\rho_{long}$) and (b) transverse ($\rho_{trans}$) resistivities measured while rotating the magnetic field in the film plane. The data are plotted as a function of the magnetic field orientation α. The lines are fits to the data using cos 2α and sin 2α functions.

Although the 180° periodicity of the oscillations is the same for AFI/Pt and FMI/Pt, we recognize striking differences in their amplitude and phase. The ADMR of the antiferromagnetic $\alpha$-$Fe_2O_3$/Pt (blue symbols in Fig. 1) and NiO/Pt (red symbols in Fig. 1) bilayers reveal the same angle dependence, which is, however, shifted by 90° relative to that of the ferrimagnetic insulator $Y_3Fe_5O_{12}$/Pt bilayer (black symbols in Fig. 1). To explain the data, we consider two antiferromagnetically coupled magnetic sublattices $\mathbf{M}^A$ and $\mathbf{M}^B$. In *ferrimagnets* with $|\mathbf{M}^A| > |\mathbf{M}^B|$ (e.g. $Y_3Fe_5O_{12}$), the net magnetization $\mathbf{M} = \mathbf{M}^A + \mathbf{M}^B$ will follow the external magnetic field $\mathbf{H}$, resulting in $\mathbf{M} \parallel \mathbf{H}$. From SMR theory [3], the resistivities then write $\rho_{\mathrm{long}}^{\mathrm{FMI}}(\alpha) = \rho_0 + \frac{\rho_1}{2}(1 + \cos 2\alpha)$ and $\rho_{\mathrm{trans}}^{\mathrm{FMI}}(\alpha) = \frac{\rho_3}{2}\sin 2\alpha$. In *antiferromagnets* with $\mathbf{M}^A = -\mathbf{M}^B$ (e.g. NiO), however, the situation is more complex. For magnetic field rotations in the easy plane, the magnetic sublattices rotate perpendicular to $\mathbf{H}$ within the film plane and we expect $\rho_{\mathrm{long}}^{\mathrm{AFI}}(\alpha) = \rho_0 + \frac{\rho_1}{2}(1 - \cos 2\alpha)$ and $\rho_{\mathrm{trans}}^{\mathrm{AFI}}(\alpha) = -\frac{\rho_3}{2}\sin 2\alpha$. Our data very nicely follow these expectations as demonstrated by fits according to the above equations (lines in Fig. 1).

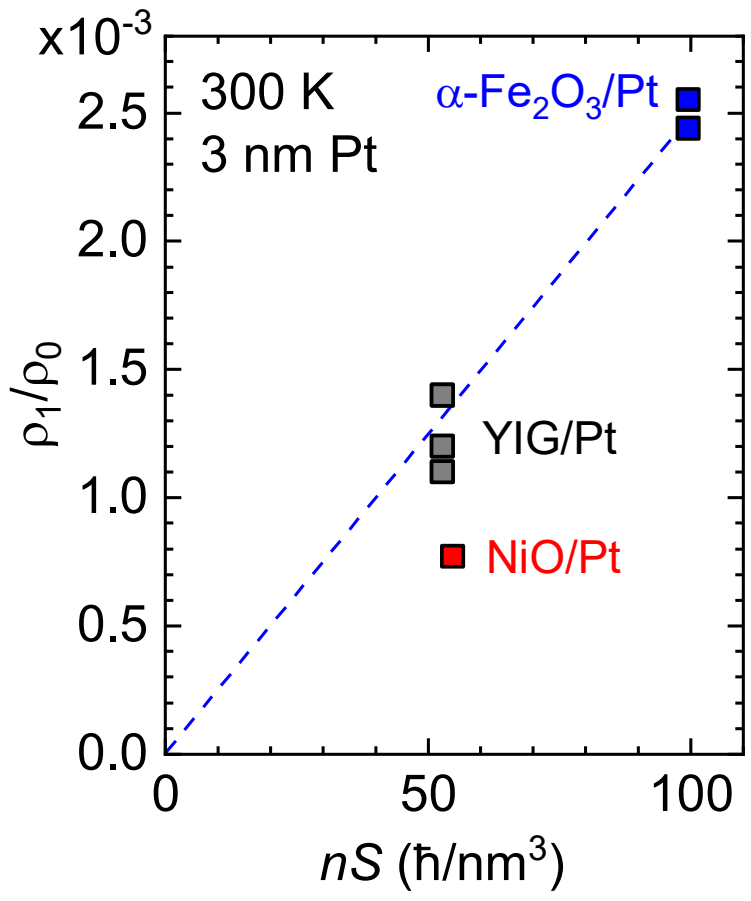

Fig. 2. Scaling behavior of the maximum SMR amplitude $\rho_1/\rho_0$ at 300 K from different FMI/Pt [2] or AFI/Pt [4,5] samples with a Pt thickness of 3 nm. The values are plotted versus the product of the spin $S$ of the magnetic ions times their volume density $n$ in the "bulk" of the FMI or AFI. The dashed blue line is a guide to the eye.

While the phase of $\rho_{\mathrm{long}}$ and $\rho_{\mathrm{trans}}$ is well understood, the amplitudes $\rho_1/\rho_0$ and $\rho_3/\rho_0$ are a matter of debate. Our epitaxial, antiferromagnetic $\alpha$-$Fe_2O_3$ thin films implemented in a heterostructure with a 3 nm thin Pt electrode show the highest SMR amplitudes reported in literature so far [4]. The exceptional value of $\rho_1/\rho_0 = 2.5 \times 10^{-3}$ even exceeds the established data from ferrimagnetic $Y_3Fe_5O_{12}$/Pt and points to a "more efficient" spin transfer between the spin polarization of the conduction electrons in the metallic Pt and the localized magnetic moments in the magnetically ordered insulator. Following indication that $\rho_1/\rho_0$ depends on the interface magnetization, we determine the volume density $n$ of the magnetic ions in the magnetically ordered materials from the dimensions and compositions of the respective unit cells. We note that we consider all magnetic ions of all sublattices, independent of their spin orientation. We find clear evidence for a monotonous correlation between $\rho_1/\rho_0$ and the $Fe^{3+}$ ion density $n$ for $Y_3Fe_5O_{12}$/Pt ($n$ = 21.10 $Fe^{3+}/nm^3$) and $\alpha$-$Fe_2O_3$/Pt ($n$ = 39.81 $Fe^{3+}/nm^3$). For NiO/Pt with its densely packed fcc structure, however, $n$ = 54.69 $Ni^{2+}/nm^3$ is highest while $\rho_1/\rho_0$ reaches only $0.77 \times 10^{-3}$ [5]. (We note, however, that in [5] $\rho_1/\rho_0$ for NiO/Pt did not saturate at the maximum available magnetic field of 17 T and is therefore underestimated.) This discrepancy is resolved by taking into account the lower spin state of $S = 1$ in $Ni^{2+}$ as compared to $S = 5/2$ in $Fe^{3+}$. Remarkably, $\rho_1/\rho_0$ follows the same universal trend when plotted versus $n$ times $S$ (Fig. 2). We note that $nS$ is equivalent to the magnetization $|\mathbf{M}|$ for *ferro*magnetic materials. In *ferri*magnets like $Y_3Fe_5O_{12}$, however, $nS$ does not represent $|\mathbf{M}|$, but the sum of the absolute values of the sublattice magnetizations $|\mathbf{M}^A| + |\mathbf{M}^B|$. The same is true for antiferromagnets with their vanishing net magnetization.

We conclude that the SMR is not an effect of the macroscopic (net) magnetization, but of all single independent magnetic sublattices in one magnetically ordered material. The SMR amplitude scales with the (volume) density of magnetic moments, multiplied with their spin state.

## III. Summary

We present a comprehensive picture of the spin Hall magnetoresistance (SMR) effect in ferrimagnetic and antiferromagnetic insulating/heavy metal thin film bilayer heterostructures. We show that the SMR provides information about the orientation of the Néel vector in AFIs. Since the SMR is a comparably simple method and also applicable at high magnetic fields, it becomes a valuable tool for reading out magnetization states in the emerging field of antiferromagnetic spintronics.

We gratefully acknowledge financial support from the German Research Foundation under Germany's Excellence Strategy – EXC-2111 – 390814868 and project AL2110/2-1.